\documentclass[twocolumn,floatfix,pra,showpacs]{revtex4-1}

\usepackage{graphicx}
\usepackage{amssymb}
\usepackage{amsmath}
\usepackage{color}
\usepackage{psfrag}
\usepackage{epsfig}
\usepackage{bbm}
\usepackage{bm}
\usepackage{simplewick} 

\newcommand{\ket}[1]{| #1 \rangle}

\newcommand{\rb}[1]{\left( #1 \right)}

\newcommand{\ew}[1]{\langle #1 \rangle}
\newcommand{\beq}{\begin{eqnarray}}
\newcommand{\eeq}{\end{eqnarray}}

\newcommand{\op}[2]{| #1 \rangle \langle #2 |}

\newcommand{\eq}[1]{Eq.~(\ref{#1})}
\newcommand{\fig}[1]{Fig.~\ref{#1}}

\newcommand{\eww}[1]{\langle\! \langle #1\rangle\! \rangle}

\begin{document}
\title{Quantum versus classical counting in nonMarkovian master equations}
\author{Clive Emary}
\affiliation{
  Institut f\"ur Theoretische Physik,
  Hardenbergstr. 36,
  TU Berlin,
  D-10623 Berlin,
  Germany
}
\author{R. Aguado}
\affiliation{
  Instituto de Ciencia de Materiales de Madrid (ICMM-CSIC),
  Cantoblanco 28049, Madrid, Spain
}

\date{\today}
\begin{abstract}
We discuss the description of full counting statistics in quantum transport with a nonMarkovian master equation.  We focus on differences arising from whether charge is considered as a classical or a quantum degree of freedom.  These differences manifest themselves in the inhomogeneous term of the master equation which describes initial correlations.  We describe the influence on current and in particular, the finite-frequency shotnoise. We illustrate these ideas by studying transport through a quantum dot and give results that include both sequential and cotunneling processes. Importantly, the noise spectra derived from the classical description are essentially frequency-independent and all quantum noise effects are absent. These effects are fully recovered when charge is considered as a quantum degree of freedom.

\end{abstract}
\maketitle

\section{Introduction}
Stochastic processes are seldom Markovian: most stochastic models arise from reduction with respect to some large collection of degrees of freedom, the evolving state of which  \emph{does} impart some memory of the system's past history. 
Despite this, most model systems are assumed to have Markovian dynamics, largely due to the inherent difficulty in treating nonMarkovian (NM) effects. If we restrict ourselves to the study of systems which can be described by a density matrix  $\rho(t)$, a generic nonMarkovian master equation (NMME) reads \cite{Nakajima-Zwanzig,Breuer}
\beq
   \frac{d}{dt} \rho(t)
   =
   \int_{t_0}^t dt'
   {\cal W}(t-t') \rho(t')
   + \gamma(t)
   \label{NMME1}
   .
\eeq
This equation describes the evolution of $\rho(t)$ under the action of NM kernel ${\cal W}(t)$ from a time $t_0$, and also includes inhomogeneous term $\gamma(t)$, which describes the effects of the systems memory of its history prior to time $t_0$.  
For the important case in which \eq{NMME1} is obtained by tracing out bath degrees of freedom, 
$\gamma(t)$ describes initial correlations between system and bath. These correlations typically decay in the long time limit and thus $\gamma(t\rightarrow\infty)\rightarrow 0$.
In a Markovian system (or Markovian approximation to the above), ${\cal W}(t-t')\sim \delta(t-t')$, memory effects are neglected and the inhomogeneity is absent.

In this paper we focus on nonMarkovian effects in electronic transport where \eq{NMME1} might, for example, describe the evolution of the density matrix of electrons in a quantum dot connected to leads.  
We are interested in the statistics of the number of charges $n$ transferred through the system in time $t-t_0$, the so-called full counting statistics (FCS) \cite{Levitov-Lesovik,noise-book}.  This information is usually encapsulated in the moment generating function (MGF), $ {\cal G}(\chi,t)$, whose derivatives give the moments 
$\langle n^m(t)\rangle\equiv\frac{\partial{\cal G}(\chi,t)}{\partial(i\chi)^m}|_{\chi=0}$.

Under a classical picture of electron transfer, \eq{NMME1} can be generalized to its $n$-resolved counterpart \cite{fli08,fli10}
\beq
  \frac{d}{dt} \rho^{(n)}(t) 
  &=&
  \sum_{n'}
  \int_{t_0}^t dt'
	  {\cal W}^{(n-n')}(t-t')\rho^{(n')}(t')
  + \gamma^{(n)}(t)
  ,
  \nonumber\\
  \label{nres1}
\eeq
which is an equation of motion for $ \rho^{(n)}(t) $, the density matrix of the system (e.g. QD) conditioned on $n$ electron transfer events having occurred.  The sum of all such partial density matrices is the full density matrix $\rho(t)=\sum_n  \rho^{(n)}(t)$. In this equation, ${\cal W}^{(n-n')}$ is that component of the kernel which transfers $n-n'$ charges, and $\gamma^{(n)}(t)$ the $n$-resolved inhomogeneity, which describes the effects of initial conditions for each value of $n$.
The probability of $n$ charge transfers is obtained by tracing over system degrees of freedom: $P(n; t)=\mathrm{Tr}\{\rho^{(n)}(t) \}$, and the MGF is simply ${\cal G}(\chi,t)=\sum_n P(n; t) e^{in\chi}$. This approach is similar to the Markovian theory \cite{Emaryetal07,Marcosetal10}.

In reality, electron number is a quantum-mechanical variable and electrons can, for example, exist in superposition between states in the dot and states in the lead where counting takes place.  The counting statistics of such a quantum degree of freedom was addressed by Levitov and coworkers by defining a MGF which yields the moments for a specific detection scheme \cite{Levitov-MGF}, an approach which has since been widely applied and has been the subject of experimental interest \cite{counting-experiments}.  

The aim of this paper is to discuss how this quantum theory of FCS can be described with an $n$-resolved NMME of the form  \eq{nres1}, and to compare and contrast quantum and classical analyses.  In the quantum treatment, \eq{nres1} arises as a reduced description from the tracing out of reservoir degrees of freedom from a fully quantum-mechanical system-reservoir theory. 
As we show, the difference between classical and quantum counting schemes
is manifest in the $n$-resolved inhomogeneity, $\gamma^{(n)}(t)$, which is uniquely determined in each case by calculating the MGF in the stationary limit.  This difference is attributed to the occurrence of system-reservoir superpositions at $t=0$, obviously absent in the classical case.
We also discuss the interpretation of $\rho^{(n)}(t)$ in the quantum case.

We note in passing that the important role of initial system-bath correlations and memory effects is, of course, not limited to $n$-resolved NMMEs. An important example is the spin-boson model \cite{Nesi07}, a prototypical model in the study of decoherence induced by external baths.

Once the above points have been clarified, we discuss how the physics at finite frequencies is affected by the classical or quantum character of $\gamma^{(n)}(t)$. In particular, we will focus on the current $\langle I(t)\rangle=\frac{d}{dt}\langle n(t)\rangle$ and its fluctuation spectrum $S(\omega)= \int_{-\infty }^{\infty }d\tau
e^{i\omega t}\frac{1}{2}\langle\{\delta {I}(t),\delta
{I}(0)\}\rangle $, where $\delta
{I}(t)={I}(t)-I_{dc}$ measures deviations away
from the steady-state current $I_{dc}=\langle {I}(t) \rangle |_{t\rightarrow\infty}$, and we have symmetrized the definition \cite{symmetry}. 
This spectrum contains information about the internal dynamics of the conductor not accessible via dc measurements
\cite{Aguado-Brandes,nag04,pil04,gala03,sal06}. Importantly, we will demonstrate that the classical description, only valid for large voltages where ${\emph n}$-resolved Markovian QMEs are justified \cite{Emaryetal07,Marcosetal10}, fails in describing frequency-resolved noise in situations where the measuring frequency is larger than both the temperature and the applied voltage, namely $\hbar\omega>k_BT,eV$. In this regime, which has been already achieved experimentally \cite{highfreqexp}, quantum noise, i.~e. zero-point fluctuations, dominates over thermal and shot noise.

As example, we will study the noise spectrum of a quantum dot with a single Zeeman-split level in the strong Coulomb Blockade regime.
In particular, we will consider the cotunneling regime using 4th-order Liouvillian perturbation theory (LPT) \cite{LPT1,ema09cot,ema11SCLPT}, which allows us to go beyond the sequential noise results of e.~g.~Refs.\cite{Engel2004,bra06,mar11}.
We show that the noise calculated from quantum counting exhibits a series of steps not only at frequencies $\hbar\omega = |\pm eV/2 - \epsilon_\sigma|$, corresponding to transitions directly between the dot levels (energy $\epsilon_\sigma$) and the leads, but also at $\hbar\omega =E_\mathrm{inelastic}$ with   $E_\mathrm{inelastic}=|\epsilon_\downarrow-\epsilon_\uparrow|$ the inelastic spin-flip transition energy. Broadening of all these quantum noise processes by elastic cotunneling is also observed.  In comparison the noise profiles derived from the classical description are extremely flat as functions of frequency and all quantum noise effects are absent.

The paper is organized as follows:  In Section \ref{sec:classical} we consider the classical NMME and derive an expression for the classical stationary inhomogeneity (\eq{gam_cl}). This expression, which was obtained previously by Flindt \emph{et al.} in Ref. [\onlinecite{fli08}], does not capture quantum noise effects.
In Section \ref{sec:quantum} where we explain how to obtain the quantum inhomogeneity  (\eq{GammaQM}) by using Liouvillian perturbation theory (technical details are relegated to the appendix). 
The current and noise spectrum obtained with our approach are described in Section \ref{sec:currentandnoise} and our results for the quantum dot example are described in Section \ref{sec:quantumdot}. 
Finally, Sec. \ref{sec:conclusions} discusses the interpretation of the quantum NMME and the relation with previous results in the literature.


\section{Classical Inhomogeneity}
\label{sec:classical}

Classically the stationary MGF for the number of charges transferred to the collector in the interval $t_0$ to $t_f$ is simply
\beq
  \mathcal{G}_\mathrm{cl}(\chi,t_{f0}) 
  &=& 
  \ew{
    e^{i \chi [n(t_f)-n(t_0)]}
  }_\mathrm{stat}
  \label{MGF_cl_defn}
\eeq
where $n(t)$ is the collector charge at time $t$.  At time $t_0$ we require that the system be stationary.  We want to calculate this quantity for a system whose dynamics are determined by a NMME.
Let $ \rho^{(n)}(t) $ be the partial system density matrix associated with presence of $n$ electrons in the collector; the sum of all such is the full density matrix of the system:  $\rho(t)=\sum_n  \rho^{(n)}(t)$.  We specify the classical NM model by stating that if the system is prepared at some initial time $t_i$ in state $\rho^{(n_i)}(t_i)$ with $n_i$ collector charges, the subsequent evolution follows the $n$-resolved NMME
\beq
  \frac{d}{dt} \rho^{(n)}(t) 
  &=&
  \sum_{n'}
  \int_{t_i}^t dt'
  {\cal W}^{(n-n')}(t-t')\rho^{(n')}(t')
  \label{NMME_from_ti}
  .
\eeq
There is no inhomogeneity here because we explicitly prepare the system without history.
Introducing the counting field $\chi$ through the Fourier transform 
$
  \rho(\chi,t) = \sum_{n} e^{i n \chi}\rho^{(n)}(t)
$, we obtain the Laplace-transform solution
\beq
  \rho(\chi,z)
  &\equiv&
  \int_0^\infty 
  d t_{fi} e^{-z t_{fi}} \rho(\chi,t_f)
  =
  \Omega(\chi,z)
  \rho(\chi,t_i) 
  ,
\eeq
with $t_{xy}=t_x-t_y$ and $\chi$-dependent propagator $\Omega(\chi,z) =\left[z-{\cal W} (\chi,z)\right]^{-1}$. 
At time $t_f>t_i$ we have then
\beq
 \rho^{(n_f)}(t_f) 
  = 
  \sum_{n_i} \Omega(n_{fi} ,t_{fi}) \rho^{(n_i)}(t_i) 
  \label{rho2}
  .
\eeq
with  $n_{xy}=n_x-n_y$ and propagator $\Omega(n ,t)$ the inverse Fourier and Laplace transform of $\Omega(\chi,z)$.  
The full system density matrix is $\rho(z) = \sum_n \rho^{(n)}(z) =  \rho(\chi=0,z)$ and, taking the long time limit, the stationary density matrix is $\rho_\mathrm{stat} = \rho(t \to \infty)= \lim_{z \to 0} z  \rho(z)$ which satisfies ${\cal W} \rho_\mathrm{stat}=0$ with ${\cal W} \equiv {\cal W}(\chi=0;z=0)$ and is assumed unique.

To calculate the MGF of \eq{MGF_cl_defn} (for arbitrary $t_{f0}$) we require not the evolution from these prepared initial conditions, but rather from initial conditions corresponding to stationarity.  This implies a history for the system and hence a finite inhomogeneity.  To this end, let us consider the system evolving under \eq{NMME_from_ti} at some intermediate time $t_0$ with $t_f > t_0 \gg t_i$.  Were the system Markovian, we could use the semi-group property of the propagator $\Omega(n ,t)$ \cite{Emaryetal07,Marcosetal10} to write \eq{rho2} equivalently as
\beq
  \rho^{(n_f)}(t_f) 
 =
  \sum_{n_0,n_i}
  \Omega(n_{f0},t_{f0}) \Omega(n_{0i} ,t_{0i}) \rho^{(n_i)}(t_i) 
  \nonumber
  .
\eeq
With a NM kernel, however, this semi-group property disappears and we therefore write the evolution from $t_i$ to $t_f$ via $t_0$ as 
\beq
  \rho^{(n_f)}(t_f) 
 & =& 
  \sum_{n_0,n_i}
  \Omega(n_{f0},t_{f0}) \Omega(n_{0i} ,t_{0i}) \rho^{(n_i)}(t_i) 
  \nonumber\\
  &&
  +
  \sum_{n_a} \int_{t_0}^{t_f} \!\!\!\!dt_a
   \Omega(n_{fa},t_{fa}) \tilde{\gamma}^{(n_a)}(t_a,t_{0i},t_i)
   ,
   \nonumber\\
  \label{nevol2}
\eeq
where the second term represents a ``correction'' to the Markovian result.  In this case $\rho^{(n)}(t) $ obeys the NMME of \eq{nres1} starting at time $t_0$ in state
$\rho^{(n_0)}(t_0) =  \sum_{n_i} \Omega(n_{0i} ,t_{0i}) \rho^{(n_i)}(t_i) $
with inhomogeneity $\tilde{\gamma}^{(n)}(t,t_{0i},t_i)$ whose arguments make explicit its dependence on previous times.
 
To obtain an explicit expression for the inhomogeneity, we equate \eq{rho2} and \eq{nevol2} and perform a Fourier transform of variable $n_f$ to obtain
\beq
  \Omega(\chi,t_{fi}) \rho(\chi, t_i) 
  &=& 
  \Omega(\chi,t_{f0}) \Omega(\chi,t_{0i}) \rho(\chi, t_i) 
  \nonumber\\
  &&\!\!\!\!\!
  +
  \int_{t_0}^{t_f} \!\!\! dt_a
   \Omega(\chi,t_{fa}) \tilde\gamma(\chi;t_a,t_{0i},t_i)
   \label{rho2chi}
\eeq
A Laplace transform over the two distinct time intervals,
$
  \int_0^\infty \int_0^\infty dt_{f0}dt_{0i} e^{-z t_{f0} - \tilde{z} t_{0i}}
$,
yields
\beq
  \frac{1}{\tilde{z}-z}
  \left\{
    \Omega(\chi, z) - \Omega(\chi, \tilde{z})
  \right\} \rho(\chi, t_i)
  ~~~~~~~~~~~~~~~~~ 
  \nonumber\\
  =
  \Omega(\chi, z)\Omega(\chi, \tilde{z}) \rho(\chi, t_i) 
  + \Omega(\chi, z)\gamma(\chi,z,\tilde{z},t_i)
  ,
\eeq
and rearranging, we obtain
\beq
  \tilde\gamma(\chi,z,\tilde{z},t_i) 
  &=&
  \frac{1}{z-\tilde{z}}
  \rb{{\cal W}(\chi,\tilde{z}) - {\cal W}(\chi,z)}
  \Omega(\chi, \tilde{z}) \rho(\chi, t_i)   \nonumber
  .
\eeq
Taking the long time limit of the $t_0-t_i$ interval and identifying the state of the system at $t_0$, the inhomogeneity reads
\beq
  \tilde\gamma^{(n)}(z) 
  &=&
  \sum_{n_0}\Gamma_\mathrm{cl}^{(n-n_0)}(z)
  \rho^{(n_0)}(t_0)
  \nonumber\\
  \Gamma_\mathrm{cl}^{(n)}(z)
  &=&
  \frac{1}{z}
   \rb{ {\cal W}^{(n)}(0) - {\cal W}^{(n)}(z) }
  \label{gamma-nonprojected}
  ,
\eeq
where the limit $t_{0i}\to \infty$ is understood here (and the state of the system at $t_i$ becomes irrelevant).
This result in \eq{nevol2} gives
\beq
  \rho^{(n_f)}(t_f) 
 & =& 
  \sum_{n_0}
    \Omega(n_{f0},t_{f0})  \rho^{(n_0)}(t_0)
  \nonumber\\
  &&
  +
  \sum_{n_a,n_0}
  \int_{t_0}^{t_f} dt_a 
  \Omega(n_{fa},t_{fa}) 
  \Gamma_\mathrm{cl}^{(n_{a0})}(t_{a0})
  \rho^{(n_0)}(t_0)
  \nonumber
  .
\eeq

The MGF for the number of charges transferred in the interval $t_f$ to $t_0$ is then
\beq
  \mathcal{G}_\mathrm{cl}(\chi,t_{f0}) 
  =
  \mathrm{Tr}
  \left\{
     \rho(\chi,t_{f0})
  \right\}
  \label{FCS_MGF_clas}
  ,
\eeq
with the Fourier-transformed density matrix
\beq
  \rho(\chi;t_{f0}) 
 & =& 
  \sum_{n_{f0}} e^{i \chi n_{f0}}
  \rho^{(n_f)}(t_f) 
  \nonumber\\
 & =& 
  \sum_{n_{f0}} e^{i \chi n_{f0}}
  \bigg\{
  \sum_{n_0}
    \Omega(n_{f0},t_{f0})  \rho^{(n_0)}(t_0)
  \nonumber\\
  &&
  +
  \sum_{n_a,n_0}
  \int_{t_0}^{t_f} dt_a 
  \Omega(n_{fa},t_{fa}) 
  \nonumber\\
  &&~~~~~~~~~
  \times
  \Gamma_\mathrm{cl}^{(n_{a0})}(t_{a0})
  \rho^{(n_0)}(t_0)
  \bigg\}
  .
\eeq
Note that the counting field $\chi$ here is associated just with the interval $t_{f0}$ and not $t_{fi}$ as in \eq{rho2chi}.
Finally, Laplace transform with respect to the interval $t_{f0}$ gives
\beq
  \rho(\chi,z)
  &=&  
  \Omega(\chi,z) 
  \left\{
    \mathbbm{1} +  \Gamma_\mathrm{cl}(\chi,z)
  \right\}
   \rho_\mathrm{stat}
   ,
   \label{rho_chiCL}
\eeq
 where in the last line we have used $\sum_{n_0}\rho^{(n_0)}(t_0) = \rho(t_0) = \rho_\mathrm{stat}$ in the long-time limit.

 The $n$-resolved density matrix $\rho^{(n)}(t)$ that corresponds to \eq{rho_chiCL} is the solution of an $n$-resolved NMME like \eq{nres1} with inhomogeneity 
\beq
 \gamma_\mathrm{cl}(\chi,z) &=& \Gamma_\mathrm{cl}(\chi; z)\rho_\mathrm{stat}
 ;
 \nonumber\\
  z\Gamma_\mathrm{cl}(\chi z)
  &=&
  {\cal W}(\chi; 0) - {\cal W}(\chi;z)
  .
  \label{gam_cl}
\eeq
Note that here $n$ is the number of charges transferred in the interval between $t$ and $t_0$ (and not the total number of collector charges).
The complete inhomogeneity is
\beq
  \gamma_\mathrm{cl}(z) 
  = \sum_n \gamma_\mathrm{cl}^{(n)}(z) = \gamma_\mathrm{cl}(\chi=0;z)
  = 
  \frac{-{\cal W}(z)}{z}\rho_\mathrm{stat}
  \label{gam0}
  ,
\eeq 
which disappears in the long time limit, $\lim_{t \to \infty }\gamma(t) = \lim_{z \to 0} z\gamma(z,t_0) = -{\cal W}(0)\rho_\mathrm{stat}=0$. 

Tracing \eq{rho_chiCL} over $n$ we obtain the MGF:
\beq
  {\cal G}_\mathrm{cl}(\chi,z)
  =
  \mathrm{Tr}
  \left\{
 \Omega(\chi,z) 
 \left[
    \mathbbm{1} +  \Gamma_\mathrm{cl}(\chi,z)
  \right]
   \rho_\mathrm{stat}
  \right\}
  \label{MGFrhogam}
  .
\eeq
With the inhomogeneity of \eq{gam0} and employing a notation in which 
 $\eww{\ldots}=\mathrm{Tr}\left\{\ldots \rho_\mathrm{stat}\right\}$ denotes the expectation value in the steady-state, the classical MGF reads
\beq
  {\cal G}_\mathrm{cl.}(\chi,z)
  =
  \frac{1}{z}
  +
  \frac{1}{z}
  \eww{
	\frac{1}{z-{\cal W}(\chi,z)}
	{\cal W}(\chi,0)
  }  
  \label{MGFCL}
  .
\eeq
The inhomogeneity of \eq{gam_cl} was given by Flindt \emph{et al} in [\onlinecite{fli08}].
and the classical MGF, like the inhomogeneity itself, can be expressed solely in terms of the $\chi$-resolved kernel.


\section{Quantum inhomogeneity}\label{sec:quantum}

The quantum-mechanical MGF of Levitov and coworkers \cite{Levitov-MGF} can be written as \cite{Bachmann2010}
\beq
  \mathcal{G}_\mathrm{qm}(\chi,t)   
  =
   \ew{
    e^{-i\frac{\chi}{2} \hat{n}}
    e^{i\chi\hat{n}(t)}
    e^{-i\frac{\chi}{2} \hat{n}} 
  }_\mathrm{stat}
  \label{FCS_MGF_lev2b} 
\eeq
where  $\hat{n}$ is the number operator of collector electrons.  
This form was originally derived by considering the precession of a spin coupled to the current flowing through the device but any ideal passive charge measurement should give the same result. 
In this section we will consider this expression in the context of a system-reservoir theory and show how the MGF may be obtained from an $n$-resolved NMME of the form \eq{nres1}. The inhomogeneity in this case is very different to that found from the foregoing classical analysis.

Let us consider a model Hamiltonian composed of reservoir, system, and interaction parts:
$
  H = H_\mathrm{res} + H_\mathrm{S} + V
$.
In its diagonal basis, the system part reads $ H_\mathrm{S} = \sum_a E_a \op{a}{a}$, where $\ket{a}$ is a many-body system state of $N_a$ electrons.
We assume noninteracting reservoirs 
$
  H_\mathrm{res} = \sum_{k,\alpha} (\omega_{k\alpha} +\mu_\alpha)
  a^\dag_{k\alpha} a_{k\alpha}
$,
where $a_{k\alpha}$ is annihilation operator for an electron of energy $\omega_{k\alpha}$ in lead $\alpha$, and $\mu_\alpha$ is the chemical potential of lead $\alpha$.
Single-electron tunnelling between system and reservoirs is described by the Hamiltonian
\beq
  V =  \sum_{k \alpha m} 
  t_{k \alpha m} a^\dag_{k\alpha} d_m 
  + t^*_{k \alpha m} d^\dag_m a_{k\alpha}
  \label{V1}
  ,
\eeq
where $d_m$ is the annihilation operator for single-particle level $m$ in the system, and $t_{k \alpha m}$ is a tunnelling amplitude. 

The density matrix $\varrho(t)$ of combined system and leads evolves according to the von Neumann equation:
\beq
  \dot{\varrho}(t) = -i \left[H,\varrho(t)\right] =  {\cal L} \varrho(t),
  \label{rhodot}
\eeq
which defines the Liouvillian super-operator ${\cal L}$.  \eq{rhodot} is solved simply as
$\varrho(t) = \Pi(t-t_0) \varrho(t_0)$ with full propagator
$\Pi(t) = e^{{\cal L} t}$.

The MGF of \eq{FCS_MGF_lev2b} can be rewritten as 
\beq
  \mathcal{G}(\chi,t) 
  &=& 
  \ew{
    e^{i H(-\chi/2)t} e^{-i H(\chi/2)t}
  }_\mathrm{stat}
  \label{FCS_MGF_lev2a}
  ,
\eeq
where the gauge-transformed Hamiltonian reads
\beq
  H(\chi) = H_\mathrm{S} + H_\mathrm{res} + V(\chi)
\eeq
with counting fields appearing in the tunnel Hamiltonian \cite{levitov-reznikov,brag06}:
\beq
  V(\chi) =  \sum_{k \alpha m} 
  t_{k \alpha m} e^{i \chi \delta_{\alpha,\gamma} } a^\dag_{k\alpha} d_m 
  + \mathrm{H.c}  
  \label{FCS_Vchi}
  ,
\eeq
where we here assume that we count only in lead $\gamma$ (hence the Kronecker delta).
The $\chi$-resolved Liouvillian super-operator is defined via
\beq
  {\cal L}(\chi) \rho = -i 
  \left\{
  H(\textstyle{\frac{1}{2}}\chi)\rho - \rho  H(-\textstyle{\frac{1}{2}}\chi)
  \right\} 
\eeq
such that, with propagator $\Pi(\chi,t) = e^{{\cal L}(\chi)t}$, the MGF of \eq{FCS_MGF_lev2a} reads ~\cite{mar11}
\beq
  \mathcal{G}_\mathrm{qm}(\chi,t_{f0}) &=& 
  \ew{
   \Pi(\chi,t_{f0})
  }_\mathrm{stat}
  \nonumber\\
  &=&
  \!\!\!\!
  \lim_{t_{0i} \to \infty}
  \!\!\!
  \mathrm{Tr} \left\{
    \Pi(\chi;t_{f0})\Pi(\chi=0;t_{0i}) \varrho(t_i)
  \right\}
  \label{FCS_LPT_MGF1}
  .
\eeq
The MGF involves two time-evolutions: first without counting such that the complete system reaches steady state at time $t_0$, and then from time $t_0$ to time $t_f$ with counting where we evolve the system with the $\chi$-dependent propagator until time $t_f$. 
For convenience we choose the initial state $\varrho(t_i)$ to be separable $\varrho(0) = \rho_0 \otimes \rho^\mathrm{res}_\mathrm{eq}$ with $\rho_0$ the initial state of the system and $ \rho^\mathrm{res}_\mathrm{eq}$ the initial state of the leads, assumed to be thermodynamic equilibrium.   In Laplace space this MGF reads
\beq
  {\cal G}_\mathrm{qm}(\chi;z)
  =
  \lim_{z_0 \to 0^+} z_0
  \mathrm{Tr} \left\{
    \Pi(\chi;z)\Pi(0;z_0) \varrho(t_0)
  \right\}
  \label{MGFQfull}
  .
\eeq

We perform the trace over the lead degrees of freedom of \eq{MGFQfull} using the technique of LPT --- the propagators are expanded as power-series of ${\cal L}_V(\chi)$ and contractions between the various tunnel vertices are then evaluated. Details of this calculation are given in appendix  \ref{appLPT} but essentially there are three possibilities: contractions solely within the leftmost propagator give rise to a $\chi$-dependent reduced system propagator at frequency $z$, $\Omega(\chi;z)$; contractions solely within the rightmost propagator give rise to the reduced system propagator at zero-frequency and without counting field;  finally, contractions {\it between} the two original propagators give rise to a block which cannot be resummed and this is the  inhomogeneous term.  After tracing out the leads and taking the long-time limit, the MGF in Laplace space reads
\beq
  {\cal G}_\mathrm{qm}(\chi;z)
  =
  \eww{
    \Omega(\chi,z)\rb{\mathbbm{1}+ \Gamma_\mathrm{qm}(\chi,z)}
  }
  ,
  \label{MGF_qm}
\eeq
which is clearly of the same form as the classical expression \eq{MGFCL} but with different individual terms.  The system propagator is
$
  \Omega(\chi,z) = [z-{\cal W}(\chi,z)]^{-1}
$
which includes the $\chi$-dependent NM effective system Liouvillian $ {\cal W}(\chi;z) = {\cal L}_\mathrm{S} + \Sigma(\chi;z)$ with self-energy $\Sigma$ arising from the coupling to the leads.  In terms of the LPT diagrams described in appendix \ref{appLPT}, to lowest order the self-energy reads (next-order diagrams are given in appendix)
\beq\label{sigma}
  \Sigma(\chi,z) =
  \contraction{}{X}{\underset{z}{-}}{X}
  X \underset{z}{-} X
  +
  \ldots
  ,
\eeq
where $X$ denotes the system part of a \emph{counting} (i. e. $\chi$-dependent ) tunneling vertex; the horizontal line, a free propagator (with frequency argument $z$); and the over-line denotes a contraction between reservoir operators. 
The stationary kernel without counting is ${\cal W} = {\cal W}(\chi=0, z=0)$ from which the stationary system density matrix is given by ${\cal W}\rho_\mathrm{stat}=0$.

In the same language, the expansion (also to lowest order) of the inhomogeneity block reads
\beq
  z\Gamma_\mathrm{qm}(\chi;z) &=&
  \rb{
  \contraction{}{X}{\underset{0}{-}}{G}
  X \underset{0}{-} G
  }
-
  \rb{
  \contraction{}{X}{\underset{z}{-}}{G}
  X \underset{z}{-} G
  }
  +\ldots
  \label{GammaQM}
  ,  
\eeq
where $G$ is the system part of a tunneling vertex without counting.
It is thus clear that the diagrams of $\Gamma_\mathrm{qm}(\chi;z)$ do indeed arise from bath contractions {\it between} two full propagations, one with and one without counting.  This has to be contrasted with the self-energy of Eq. (\ref{sigma}) in which \emph{all} vertices are $\chi$-dependent.  This emphasizes that the quantum inhomogeneity, and by extension the MGF cannot, in general, be couched solely in terms of $\Sigma(\chi,z)$.  This is in contrast with the classical case where knowledge of the kernel ${\cal W}(\chi;z)$ is sufficient to determine the inhomogeneity and the MGF.

The moment generating function can be written as the trace over the $\chi$-resolved density matrix 
\beq
  \rho(\chi,z)=
  \Omega(\chi,z)
  \left[
    \mathbbm{1}+ \Gamma_\mathrm{qm}(\chi,z)
  \right] 
  \rho_\mathrm{stat}
  .
\eeq
As in the classical case the corresponding $n$-resolved density matrix is the solution of an NMME of the form \eq{nres1} with inhomogeneity:
\beq
  \gamma(\chi;z) = \Gamma_\mathrm{qm}(\chi,z)\rho_\mathrm{stat}
  .
\eeq
We refer to the discussions section concerning the interpretation of these results in terms of an $n$-resolved NMME.
Setting $\chi =0$ into expression \eq{GammaQM} gives the full quantum inhomogeneity to be
\beq
  \gamma_\mathrm{qm}(z) 
  &=& \Gamma_\mathrm{qm}(0;z) \rho_\mathrm{stat}
  =- {\cal W}(z) \rho_\mathrm{stat}
\eeq
which is identical to the expression found in the classical case.

\section{Current and Noise}\label{sec:currentandnoise}
We now derive expressions for the current and noise based on our two results for the MGF, classical (\eq{MGFCL}) and quantum (\eq{MGF_qm}).  The average stationary current is defined as the rate-of-change in the number of collector electrons: $\ew{I(t)} = \frac{d}{dt} \ew{n(t)}$
In Laplace space we have
\beq
  \ew{I (z)}
  &=& 
  z \ew{n(z)}
=
   z
   \left.
   \frac{\partial }{\partial (i\chi)} {\cal G}(\chi,z)
   \right|_{\chi \to 0}
\eeq
The corresponding finite-frequency noise can be quantified with correlation function
\beq
  S(\omega) 
  \equiv
   \int_{-\infty}^\infty dt
   e^{i \omega t}
   \frac{1}{2}\ew{\left\{\delta I(t),\delta I (0)\right\}}
\eeq
with $\delta I = I - \ew{I}$ and where we have symmetrised the definition \cite{symmetry}. 
Using MacDonald's formula \cite{MacDonald} this can be expressed as
\beq
  S(\omega) 
  &=& 
  \omega \int_{-\infty}^\infty dt
  \sin( \omega t)
  \frac{d}{dt}
  \left[
    \ew{n^2(t)} - \ew{n(t)}^2
  \right]
\eeq
and, following Ref. [\onlinecite{fli08}], as
\beq
  S (\omega) =
  - \frac{\omega^2}{2} \frac{\partial^2}{\partial (i\chi)^2}
  \left\{
    \ew{{\cal G}(\chi,z=i\omega) }
   +(z=-i\omega)
  \right\}_{\chi \to 0}
  \nonumber
  .
\eeq

In discussing results, we will consider the total noise, rather than that in a single contact.  According to the Ramo-Shockley theorem, for a two-terminal conductor the total current can be written in terms of the currents through the left and right terminals as: $I_{\mathrm{tot}}=\alpha I_L + \beta I_R$.  This change can be affected by including counting fields for both contacts ($\chi_L$ and $\chi_R$ for left and right leads respectively) and making the substitutions
 $\chi_{\mathrm{tot}}\equiv\chi_L+\chi_R$, and $\chi_{\mathrm{accum}}\equiv \beta\chi_L-\alpha\chi_R$.  Differentiation of the MGF with respect to 
 $\chi_{\mathrm{tot}}$ then give the total cumulants rather than those of a single lead \cite{Marcosetal10}. In the following we consider symmetric barriers and set $\alpha=\beta=1/2$.

%
We begin with the classical case and exapnd the kernel as
\beq
  {\cal W}(\chi,z) = \sum_{k=0} \frac{(i \chi )^k}{k!} {\cal W}^{(k)}(z)
  ,
\eeq
with 
\beq
  {\cal W}^{(k)}(z) 
  = 
  \left.
    \frac{\partial^k}{\partial (i \chi)^k}  {\cal W}(\chi,z)
  \right|_{\chi \to 0}
  .
\eeq
The component ${\cal W}^{(0)}(z) = {\cal W}(z)$ is the non-counting kernel and
$\Omega(z) = \left[z-{\cal W}^{(0)}(z)\right]^{-1}$ is the standard propagator without counting field.

Differentiating the classical MGF of \eq{MGFCL}, we obtain
the current
$
  \ew{I}_\mathrm{cl.}(z) = z \eww{{\cal W}^{(1)}(0)}
$,
such that the current
\beq
  \ew{I}_\mathrm{cl.}(t) = \ew{I}_\mathrm{cl.} = \eww{{\cal W}^{(1)}(0)}
  ,
\eeq
is stationary and depends only on the first derivative block at zero-frequency.  Similarly, the noise reads
\beq
  S_\mathrm{cl.}(\omega)
  &=&
  \eww{ 
    {\cal W}^{(2)}(0)
  }
  +
  \eww{
    {\cal W}^{(1)}(i\omega) \Omega(i\omega) {\cal W}^{(1)}(0)
  } 
  \nonumber\\
  &&
  +
  \eww{
   {\cal W}^{(1)}(-i\omega) \Omega(-i\omega) {\cal W}^{(1)}(0)
  }
  \label{SCL}.
\eeq
The first term is the autocorrelation term and is evaluated at zero frequency.  The remaining terms contain first-derivative blocks evaluated at both zero and finite-frequency.

%
Taking the first derivative of the quantum MGF of \eq{MGF_qm}, rearranging and canceling, we find the stationary current
\beq
  \ew{I}_\mathrm{qm.} =  \eww{{\cal J}^{(1)}(\cdot,0)}.
  \label{Iqm}
\eeq
Here the current block ${\cal J}^{(1)}$ (to sequential order) is
\beq
  {\cal J}^{(1)}(z_1,z_0) &=& 
  \contraction{}{G}{\underset{z_1}{-}}{X'}
  G \underset{z_1}{-} X'
  +
   \contraction{}{X'}{\underset{z_0}{-}}{G}
  X' \underset{z_0}{-} G
  + \ldots
  \label{J1seq}
  ,
\eeq
where $X'$ is the derivative of the system part of the $\chi$-dependent vertex evaluated at $\chi= 0$.
As it appears in \eq{Iqm}, the leftmost frequency argument of this block is replaced with a dot since all diagrams with leftmost $G$-superoperator, and thus all those containing $z_1$, are exactly zero in the stationary expectation value \cite{LMGFN}.

Similarly, taking the second derivative and simplifying we find the quantum expression for the noise reads
\beq
  S_\mathrm{qm}(\omega) &=& \eww{  
  \textstyle{\frac{1}{2}}{\cal J}^{(2)}(\cdot,i\omega,0)
  +
  {\cal J}^{(1)}(\cdot,i\omega) \Omega_\mathrm{S}(i\omega){\cal J}^{(1)}(i\omega,0)} 
  \nonumber\\
  +
  &&
  (\omega\rightarrow-\omega),
  \label{SQM}
\eeq
with ${\cal J}^{(1)}$ block as as in \eq{J1seq} and with second-order block
\beq
 {\cal J}^{(2)}(z_2,z_1,z_0) 
  &=& 
  2 \contraction{}{X'}{\underset{z_1}{-}}{X'}
  X' \underset{z_1}{-} X'
  +  
  \contraction{}{X''}{\underset{z_0}{-}}{G}
  X'' \underset{z_0}{-} G
  +
  \contraction{}{G}{\underset{z_2}{-}}{X''}
  G \underset{z_2}{-} X''
  + \ldots
  .
  \nonumber
\eeq
Expression \eq{SQM} reproduces the sequential expressions of Refs.~\cite{Engel2004,bra06, mar11} and, as is made explicit in the appendix, extends them to cotunneling and in principle beyond.

%
It is clear that the classical and quantum expressions for both the current and the noise have the same general block structure.  The blocks themselves differ between the two approaches, however.  In the classical case, one has the simple ${\cal W}$-derivative blocks, whereas in quantum case, the ${\cal J}$-blocks are not directly related to the original kernel and must be calculated separately.
We can directly compare classical and quantum results if we consider the classical Liouvillian ${\cal W}(\chi;z)$ to be that arising from our microscopic quantum model --- the only difference between the two calculations is then the nature of the counting and hence the inhomogeneity.
In this case the blocks $\mathcal{W}^{(k)}$ and $\mathcal{J}^{(k)}$ have the same structure, but differ in the assignment of free-propagator frequencies within the diagrams. For example, the first-derivative block ${\cal W}^{(1)}(z)$ has the diagrammatic expansion
\beq
  {\cal W}^{(1)}(z) &=& 
  \contraction{}{G}{\underset{z}{-}}{X'}
  G \underset{z}{-} X'
  +
   \contraction{}{X'}{\underset{z_0}{-}}{G}
  X' \underset{z}{-} G
  + \ldots
  \label{W1seq}
  ,
\eeq
This is the same as the ${\cal J}^{(1)}$ block, \eq{J1seq} but for the frequency arguments.

Since, in both expressions for the stationary current, zero is the only relevant frequency argument, the two currents are equal
\beq
  \ew{I}_\mathrm{cl} =  \ew{I}_\mathrm{qm}
  .
\eeq
This is as we expect since memory effects should be irrelevant for the stationary properties of the system.
In fact, since the inhomogeneity vanishes in the $z\to 0$ limit, the zero-frequency FCS of classical and quantum approaches must be identical.  We find the zero-frequency limit of the noise
\beq
  S_\mathrm{qm.}(0) 
  &=& 
  S_\mathrm{cl.}(0) 
  \nonumber\\
  &=&
  \eww{ 
    {\cal W}^{(2)}
  }
  + 2
  \eww{
    {\cal W}^{(1)} \mathrm{R}{\cal W}^{(1)}
  } 
  \nonumber\\
  &&
  + 
  2 \ew{I}
  \rb{
    \eww{\dot{{\cal W}}^{(1)}} +  \eww{{\cal W}^{(1)} \mathrm{R}\dot{{\cal W}}}
  }
  \label{S02}
  ,
\eeq
with all blocks evaluated as zero-frequency, $\dot{X} = \partial X / \partial z$, 
and $\mathrm{R}(z)$ the projection of the propagator $\Omega(z)$ out of the stationary state. 
which (up to definitions) agrees Ref.~\cite{fli08,flithesis}.
At finite frequency, however, the MGFs and the noise results from the two approaches do differ.  We explore these differences by considering an example from quantum transport in the next section.

\section{Quantum transport results}\label{sec:quantumdot}

We consider the transport through a quantum dot with a single Zeeman-split level that we model with the Anderson Hamiltonian
\beq
  H = 
   \sum_\sigma \epsilon_\sigma d_\sigma^\dagger d_\sigma
  + U n_\uparrow n_\downarrow
  + H_\mathrm{res}
  + V,
\eeq
with $\epsilon_\sigma$ the energy of a spin-$\sigma$ electron in the dot and $U$ the interaction energy.  The reservoir and interaction Hamiltonians are as section III, with index $\alpha$ including both lead ($=L,R$) and spin ($=\uparrow,\downarrow$) index.  For simplicity we take the limit of large Coulomb repulsion, $U \to \infty$, and thus restrict the Hilbert space of the QD to three states: ``empty'', and spin-up and spin-down single-electron states.
We consider a bias symmetric about zero, $\mu_L=-\mu_R = \frac{1}{2} e V$, and two different level configurations.  In both cases, we discuss the total noise  $S(\omega)$.

\begin{figure}[tb]
  \begin{center}
  \epsfig{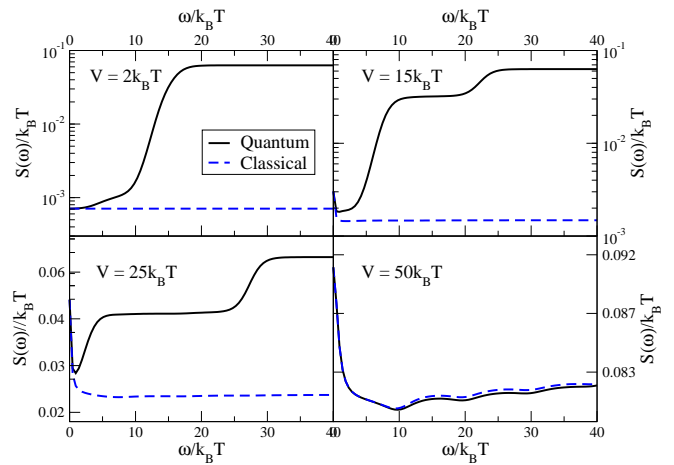}
  \caption{    
    The finite-frequency total noise $S(\omega)$ as a function of $\omega$ for the infinite-$U$ Anderson model QD with level positions $\epsilon_\uparrow=-15 k_B T$, $\epsilon_\downarrow=5 k_B T$ and chemical potentials $\mu_L=-\mu_R = \frac{1}{2} e V$; $V/k_BT = 2,15,25,50 $.
    Results are shown for the noise calculated with both quantum  and classical counting from fourth-order (cotunneling) LPT kernels (solid black and dashed blue lines respectively).
    Both counting theories give the same noise at $\omega=0$ but as frequency increases, differences start to develop.
    This is most evident at small biases where, whereas the classical noise is essentially flat as a function of frequency, the quantum result shows marked steps at $\omega=|\mu_\alpha - \epsilon_\uparrow|$ (steps at $\omega=|\mu_\alpha - \epsilon_\downarrow|$ are also visible at $V=50k_BT$; see text). For $V=2k_BT$, the difference between between the two results is approximately two orders-of-magnitude at high frequencies.
    At higher bias the system enters the shotnoise regime and the relative importance of the quantum contributions lessens.
    Other parameters were $\Gamma_L=\Gamma_R= \frac{1}{4} k_B T$ and bandwidth $X_C=300 k_B T$.
    \label{FIG_Sw}
 }
  \end{center}
\end{figure}

\begin{figure}[tb]
  \begin{center}
  \epsfig{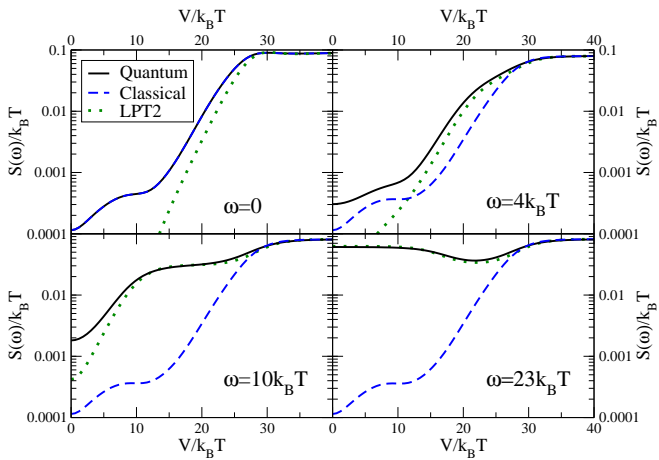}
  \caption{
    As \fig{FIG_Sw} but with finite-frequency noise $S(\omega)$ plotted as a function of applied bias for $\omega/k_BT =0,4,10,23$.     
    Quantum- and classical-counting results with fourth-order LPT (cotunneling) kernels (black solid and blue dashed, respectively) are show alongside results from quantum counting  with second-order LPT (sequential) kernels (green dotted lines).
    The quantum results are dominated by sequential processes with shotnoise step at $V = 30 k_BT$ and quantum noise step up at $V = 30 k_BT -2\omega$ ($\omega < 15k_BT$) and step down at $V = -30 k_BT +2\omega$ for high frequencies ($\omega > 15k_BT$)  instead.   With the parameters cotunnellng predominantly serves to increase noise, both at zero and finite frequency, at low frequencies.
    In contrast to the quantum case, the classical results change very little with increasing frequency
    \label{FIG_SeV}
 }
  \end{center}
\end{figure}

In the first configuration we set $\epsilon_\uparrow=-15 k_B T$, $\epsilon_\downarrow=5 k_B T$ and sweep bias and frequency.
Results for this configuration are shown in \fig{FIG_Sw} and \fig{FIG_SeV} as calculated using 4th-order LPT kernels (which include sequential and cotunneling contributions) and, in the case of \fig{FIG_SeV}, quantum results calculated with 2nd-order LPT kernels (sequential processes only).
At zero frequency, classical and quantum results coincide as expected. At finite frequencies, the contrast is stark.
Concentrating on \fig{FIG_Sw} first, we observe that, for low bias, the classical expression predicts a noise that is essentially flat as a function of frequency.  In contrast, the quantum noise increases in a step-wise fashion. These steps are the quantum-noise steps described in Ref.~[\onlinecite{mar11}] arising from quantum fluctuations between system and reservoir.  With sequential processes dominant, we expect quantum noise steps at $\omega = |\mu_\alpha - \epsilon_\sigma|$ for some $\alpha=L,R$ and $\sigma=\uparrow,\downarrow$. For $V\lesssim 30 k_BT$ only the transition from the $\epsilon_\uparrow$ is visible since, due to the Coulomb blockade, this is the only level populated in the steady state.  For $V=50k_BT$ both levels lie within the transport window and all four transitions are visible.

\fig{FIG_SeV} shows the noise as a function of bias. At a bias of $V = 2\epsilon_\uparrow$ the lower-lying level enters the bias window resulting in a step as we enter the shotnoise regime.  Noise results at and above this step are insensitive to calculational details since sequential electron hopping, which is essentially classical in the weakly coupled limit, then dominates.  Below this voltage, quantum noise effects are visible.
The dominant quantum noise process here is the excitation of an electron from the lower state (energy $\epsilon_\uparrow$) to the leads. This is a sequential process and can occur for a bias $V < 2( \epsilon_\uparrow + \omega)$ to the left lead ($\mu_L=+V/2$),
and for $V > -2 (\epsilon_\uparrow + \omega)$ to the right lead ($\mu_R=-V/2$).  With $V>0$ and for $\omega < |\epsilon_\uparrow|$, we therefore obtain a single step up at $V=-2 (\epsilon_\uparrow+\omega)$.  For $\omega > |\epsilon_\uparrow|$, exitation to both leads is possible at $V=0$, and low bias noise is correspondingly high.  At $V=2(\epsilon_\uparrow+\omega)$ one of these channels closes and we obtain a step down.

Considering now both sequential and cotunneling processes together, the zero-frequency curve of \fig{FIG_SeV} reproduces the results of Ref.~[\onlinecite{ema09cot}] and, of course, classical- and quantum-counting results agree. The most significant new feature that occurs at cotunneling order is a broad feature at low bias --- following Ref.~\cite{thielmann} we associate this with inelastic cotunneling centered about the  spin-flip transition energy 
$E_\mathrm{inelastic}=|\epsilon_\downarrow-\epsilon_\uparrow|$.
For these parameters, the main extra contribution of cotunneling processes at finite frequencies is to increase the fluctuations at very small bias.  Quantum noise features directly attributable to quantum cotunneling are masked by sequential processes here. 
The classical results at finite frequency differ strongly from the quantum case --- at finite frequencies, the basic form of the classical noise is broadly the same as at zero frequency (sequential step at $V=2|\epsilon_\uparrow|$, flat inelastic cotunneling feature at low bias). There is a slight shift downwards of the classical curve for $\omega>0$ compared with $\omega=0$, but this is a relatively small effect.

The finite-frequency Fano factor $F(\omega) = S(\omega) /\ew{I}$ further serves to exaggerate the effects of quantum noise (not shown).  At zero frequency, the two main features of the Fano factor as a function of voltage are a divergence of the Fano factor for $e V \to 0$, consistent with the fluctuation-dissipation theorem, and a pronounced superPoissonian region strongly influenced by inelastic spin-flip cotunneling processes. \cite{thielmann,ema09cot}.
In  the classical case finite-frequency measurement serves to reduce the Fano factor such that noise becomes subPoissonian everywhere except for the divergence at $V \to 0$.  In the quantum case, however, the superPoisonian peak increases in height and moves to lower biases.  The result is massively increased Fano factor ($ F(\omega)> 100$) for a significant bias range $V \lesssim E_\mathrm{inelastic}$.

\begin{figure}[tb]
  \begin{center}
  \epsfig{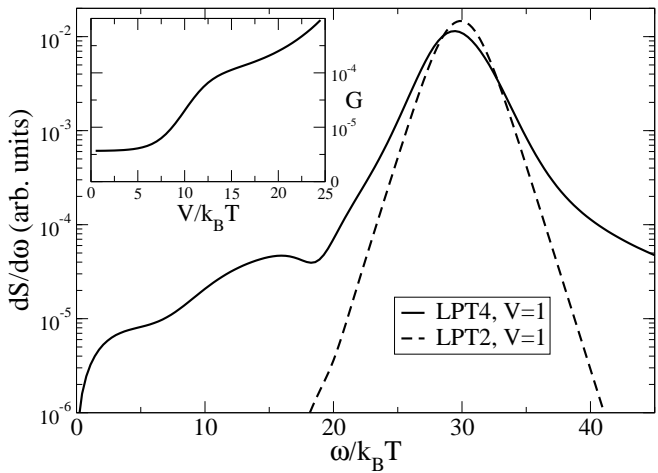}
  \caption{
    Transport through the infinite-$U$ Anderson model with both dots levels well below transport window: $\epsilon_\uparrow=-20 k_B T$, $\epsilon_\downarrow=-30 k_B T$; other parameters as in \fig{FIG_Sw}.
    The inset shows the differential conductance $G=d\ew{I}/d(V)$ as a function of bias. A clear jump is observed around the inelastic cotunneling energy $E_\mathrm{inelastic}=|\epsilon_\downarrow-\epsilon_\uparrow| = 10k_B T$.
    The main panel compares the derivative of the shotnoise, $dS(\omega)/d\omega$, as a function of $\omega$ for both sequential (LPT2) and cotunneling (LPT4) kernels near linear response ($V=1k_BT$).  Both predict a large peak when the measurement frequency is approximately equal to $\mu_{L/R}-\epsilon_\downarrow\sim 30 k_BT$.
    In the LPT4 results, this peak is broadened by elastic cotunneling processes, but more importantly, there is a significant enhancement in the low-bias tail which can be attributed to resonance with inelastic cotunneling process $\omega \sim E_\mathrm{inelastic}$.
    \label{FIGinelastic}
 }
  \end{center}
\end{figure}

Figure \ref{FIGinelastic} shows results for a configuration in which both dot levels are far below the transport window ($\epsilon_\uparrow=-20 k_B T$, $\epsilon_\downarrow=-30 k_B T$) such that d.c. transport can only proceed by cotunneling processes.  The differential conductance $G=d\ew{I}/d(V)$ is plotted in the inset which would be vanishingly small with just sequential terms in the kernel, but is finite with cotunneling terms included here.  Of particular note is a step in the conductance at the inelastic cotunneling energy $V \sim E_\mathrm{inelastic}$ \cite{Franceschi}.

The main panel of \fig{FIGinelastic} shows the derivative of the quantum shotnoise $dS_\mathrm{qm}(\omega)/d\omega$ as a function of $\omega$.  Results are plotted from both second- and fourth-order kernels, labelled LPT2 and LPT4 respectively.
The dominant feature in this plot is a large peak centered at a frequency $\omega$ resonant with the transition between the lowest dot state and the chemical potentials, i.e. for $\omega = |\mu_\alpha - \epsilon_\downarrow|$. Since we are here near linear response,
$\mu_L\approx \mu_R$,  we have a single peak --- increasing the bias splits this peak in two (not shown).

As comparison of the LPT2 and LPT4 results show, cotunneling processes modify this sequential peak in two ways. Elastic cotunneling processes serve to broaden the sequential resonance.  More importantly, cotunneling leads to a significant increase in $dS(\omega)/d\omega$ on the low-frequency side of the main resonance.  This we attribute to the effects of inelastic cotunneling and this feature is situated around resonance with the inelastic cotunneling energy $\omega =E_\mathrm{inelastic}$.  We note that this feature is rather broad and does not have a simple peak structure.
For comparison, the classical $dS(\omega)/d\omega$ is several orders of magnitude smaller than the quantum result (too small to plot on the same scale as \fig{FIGinelastic}), and is approximately flat as a function of $\omega$ and shows none of these resonant features.

\section{Discussion}\label{sec:conclusions} 

We have discussed how the description of the FCS of a transport system in terms of an $n$-resolved NMME differs depending on whether the number of transferred electrons is considered a classical or a quantum variable.  
The primary difference manifest is the inhomogeneity --- in the quantum case, the inhomogeneity can not be described in terms of the system Liouvillian alone but rather requires new diagrams which reflect initial superpositions between system and reservoir that are obviously absent in the classical case. 
This difference is not observable in the long-time properties of the system, in particular the zero-frequency FCS, since $\gamma(\chi, t\to \infty)=0$.  However, as we have discussed in general and for our quantum dot example, this difference is visible in the finite-frequency FCS and the influence on high-frequency quantum noise can be dramatic.

The interpretation of the $n$-resolved NMME and the associated probability $P(n,t)$ in the quantum case is interesting.  As is clear from \eq{GammaQM}, the inhomogeneity contains counting field factors $e^{i\chi(2k+1/2)};k=0,\pm1,\pm2,\ldots$ due to the pairing of counting ($X$) and non-counting ($G$) vertices (in the self-energy, counting vertices are always paired with other counting vertices).
The presence of non-integer powers of $e^{i\chi}$ means that for the $n$-resolved NMME to contain the same information as its $\chi$-resolved counterpart, the transformation from $\chi$ to $n$ representation should be a continuous Fourier transform (cf. the classical case, where the $n$ is clearly integer and the discrete transform is appropriate) and the above counting field factors imply an interpretation in terms of half-integer charge.  Furthermore, in the quantum case there is no guarantee that $P(n,t)$ is positive; The best interpretation of $P(n,t)$ and, by extension, the diagonal elements of $\rho^{(n)}(t)$, is as quasi-probabilities \cite{Bednorz2008,interpFN}.

Interestingly, Shelankov and Rammer \cite{Shelankov-Rammer} introduced the MGF
\beq
  \mathcal{G}_\mathrm{SR}(\chi,t) 
  =
  \sum_n
  \ew{
    \hat{P}_n
    e^{i H(-\chi/2)t} e^{-i H(\chi/2)t}
    \hat{P}_n
  }
  \label{FCS_MGF_shel}
 ,
\eeq
with $\hat{P}_n$ being a projection operator onto states of definite charge $n$ (a similar MGF was considered by Esposito et al. \cite{Esposito}) that removes the interpretational problems of the MGF in \eq{FCS_MGF_lev2a}. This is achieved by removing all traces of charge superposition in the initial density matrix.  At sequential order, this projection approach is equivalent to our classical approach; at higher orders the approaches differ since \eq{FCS_MGF_shel} still admits superpositions within a charge-sector, which are absent in our classical approach. Nevertheless, since such superpositions do not contribute to the charge fluctuations at $t_0$, we anticipate very similar results for \eq {FCS_MGF_shel} and our classical approach.

\acknowledgements{
We are grateful to C.~Flindt, A.~Braggio, and T.~Brandes for useful discussions.
Financial support was provided by DFG projects BR 1528/8-1
 and SFB 910, and by MICINN Spain grant FIS2009-08744.
}


\appendix

\section{Liouvillian Perturbation theory \label{appLPT}}

For completeness, we review here the basics of LPT and the translation of the resulting kernels into diagrams.  More details are to be found in Refs.~\cite{LPT1,ema09cot} and \cite{ema11SCLPT} for the diagrams.

To ease book-keeping, we introduce a compact single index ``$1$'' to denote the triple of bath indices $(\xi_1,k_1,\alpha_1)$.  The first index $\xi_1=\pm$ indicates whether a reservoir operator is a creation or annihilation operator:
\beq
  a_1 =  a_{\xi_1 k_1 \alpha_1}
  =
  \left\{ 
    \begin{array}{c c}
      a^\dag_{k_1\alpha_1}, &\quad \xi_1 =+\\
      a_{k_1\alpha_1}, &\quad \xi_1 =-
    \end{array}
  \right. 
  .
\eeq
Leaving sums implicit, the reservoir Hamiltonian reads
\beq
  H_\mathrm{res} &=&\rb{ \omega_{k \alpha}+\mu_\alpha} a_{+k\alpha} a_{-k\alpha}
  \nonumber \\
  &=& 
  \rb{ \omega_1+\mu_1} a_1 a_{\overline{1}}\delta_{\xi_1,+}
  \label{AP_Hres},
\eeq
where $\overline{1}$ denotes $(-\xi_1,k_1,\alpha_1)$.
In equilibrium, the reservoir electrons are distributed according to the Fermi function
$f(\omega) = \rb{e^{\omega/k_B T} +1}^{-1}$,
which, since we include the chemical potential in \eq{AP_Hres} and assume a uniform temperature, is the same for all reservoirs.

We then define the system operators
$
  g_{k\alpha} 
  =
  \sum_{m}  t_{k \alpha m} j_{m}
  \label{gtj}
$,
such that the interaction Hamiltonian of \eq{V1} can be written
$V = \xi_1 a_1 g_1$.
Correspondingly, the tunnel Liouvillian reads
\beq
  {\cal L}_V 
  = -i \left[V,\cdot\right]
  =-i \xi_1 \sum_p p \sigma^p A^p_1 G^p_1
  \label{LVdef}
\eeq
with bath-superoperator $A$ defined as 
\beq
  A^p_1 O = 
  \left\{
   \begin{array}{c c}
      a_1 O, &p =+ \\
     O a_1, &p =-
    \end{array}
  \right.
  \label{ddef}
  ,
\eeq 
and system superoperator $G$ via 
\beq
  G^p_1 O = 
  \sigma^p 
  \times
  \left\{
   \begin{array}{c c}
      g_1 O, &p =+ \\
      - O g_1, &p =-
    \end{array}
  \right.
  \label{Gdef}
  .
\eeq
The object $\sigma^p$ is a dot-space superoperator with matrix elements
\beq
  \rb{\sigma^p}_{ss',\bar{s}\bar{s}'} = 
  \delta_{s\bar{s}}\delta_{s'\bar{s}'} 
  \left\{
   \begin{array}{c c}
      1,\quad N_s - N_{s'}=~\mathrm{even} \\
      p,\quad N_s-N_{s'}=~\mathrm{odd}
    \end{array}
  \right.
  ,
\eeq
where, $N_s$ is the number of electrons in state $s$.

The von Neumann equation reads $\dot \varrho = {\cal L}\varrho$, with Liouvillian super-operator
\beq
  {\cal L} = {\cal L}_\mathrm{res} + {\cal L}_\mathrm{S} + {\cal L}_V
\eeq
with ${\cal L}_\mathrm{res}= -i\left[H_\mathrm{res},\bullet~\right]$,  ${\cal L}_\mathrm{S}=-i\left[H_\mathrm{S},\bullet~\right]$, and 
$ {\cal L}_V = -i\left[V,\bullet~\right]$.
The solution in Laplace space is $ \varrho(z) =\left[z-{\cal L}\right]^{-1}  \varrho(0) $.
The reduced density matrix of the dot is given by tracing out the electron reservoirs
\beq
  \rho_\mathrm{S}(z) 
  = 
  \mathrm{Tr}_\mathrm{R} \left\{\frac{}{} \rho(z) \right\}
  = 
  \mathrm{Tr}_\mathrm{R}  
  \left\{\frac{}{}
    \frac{1}{z-{\cal L}}  \rho(0)
  \right\}
  .
\eeq
This we expand in powers of ${\cal L}_V$ to obtain
\beq
  \rho_\mathrm{S}(z) 
  =  
  \mathrm{Tr}_\mathrm{R}  
  \left\{
    \rb{
      \Pi_0(z)
      +
      \Pi_0(z){\cal L}_V\Pi_0(z)
      +\ldots
    }
    \rho(0)
  \right\}
  \label{rhoD_exp}
\eeq
with free propagator
$
  \Pi_0(z) = \left[z-{\cal L}_\mathrm{res}-{\cal L}_\mathrm{S}\right]^{-1}$.
With substitution of \eq{LVdef}, a typical term of the expansion  \eq{rhoD_exp} reads
\beq 
  (-i)^n
  \rb{\prod_{l=1}^n \xi_l p_l}
  \mathrm{Tr}_\mathrm{R}  
  \left\{\frac{}{}
    \Pi_0(z)
    \sigma^{p_n}  A^{p_n}_n G^{p_n}_n
    \ldots
    \right.
    \nonumber\\
    \ldots
    \left.
    \sigma^{p_1} A^{p_1}_1 G^{p_1}_1
    \Pi_0(z)
    \rho(0)
    \frac{}{}
  \right\}
  .
\eeq
Evaluating the action of the $\sigma^p$ superoperators we obtain a factor $\prod_{l}^{\mathrm{odd}} p_l$ and, as the $G$ operators also contain $\sigma$, they evaluate at different positions in the chain as
\beq
  G^{p_{l}}_{l}O 
  &=& 
  (p_l)^{l+1}
  \times
  \left\{
   \begin{array}{c c}
      g_l O , &p_l =+\\
      O g_l, &p_l =-
    \end{array}
  \right.
  .
\eeq
We then bring all the $A$ operators to the right of the $G$ operators by using the relations
$
  A_1^p G_{1'}^{p'} = - p p'  G_{1'}^{p'} A_1^p
$
,
$
  \mathrm{Tr}_\mathrm{R} {\cal L}_\mathrm{res} 
  = 0
$
,
$  
  {\cal L}_\mathrm{res} \rho_\mathrm{res}^\mathrm{eq} =0
$
and
$
  A_1^p  {\cal L}_\mathrm{res} = \rb{  {\cal L}_\mathrm{res} - x_1 }A_1^p 
$
with
\beq
  x_1 &=& -i \xi_1 (\omega_1 + \mu_{\alpha_1})
.
\eeq
and our typical term becomes
\beq
  &&
  (-i)^n \rb{\prod_{l'}^{\mathrm{odd}} p_{l'}}
  \mathrm{Tr}_\mathrm{R}  
  \left\{\frac{}{}
    \Omega_0(z)
    G^{p_n}_n 
    \Omega_0(z_{n-1})
    G^{p_{n-1}}_{n-1} 
    \ldots
  \right.
  \nonumber\\
  &&
  \left.
    \ldots 
     G^{p_2}_2 
    \Omega_0(z_1)
    G^{p_1}_1 
    \Omega_0(z)
    A^{p_n}_n
   \ldots
    A^{p_1}_1
    \rho(0)
  \right\}
  \label{typ_X}
\eeq
with free dot propagator
\beq
  \Omega_0(z) = \frac{1}{z-{\cal L}_\mathrm{S}}
  ,
\eeq
and 
$
  z_m = z+ \sum_{l=m+1}^{n} x_l ;
  \quad
  1 \le m \le n-1
$.

The the reservoir expectation values,
$
  \mathrm{Tr}_\mathrm{res}  
  \left\{\frac{}{}
    A^{p_n}_n
   \ldots
    A^{p_1}_1
    \rho_\mathrm{res}^\mathrm{eq}
  \right\}
  =
  \ew{A^{p_n}_n
   \ldots
    A^{p_1}_1
    }_\mathrm{eq}
$, can be evaluated with the rules of Wick's theorem in Liouville space, which read \cite{LPT1}
\begin{itemize}
  \item decompose $\ew{\ldots}_\mathrm{res}$ into pair-contractions
  \item add minus sign for each interchange of $A$
  \item omit factor $ \rb{\prod_{l'}^{\mathrm{odd}} p_{l'}}$ arising from 
    $\sigma$ super-operators
  \item each pair contraction then contributes a factor
  \beq
   \ew{A_2^{p_2}A_1^{p_1}}=
  \gamma_{21}^{p_2p_1}=\delta_{2\overline{1}} p_1 f(-\xi_1 p_1 \omega_1)
  .
  \eeq
\end{itemize}
With these rules, our typical term becomes
\beq
  (-i)^n 
  \Omega_0(z_n)
  G^{p_n}_n 
  \Omega_0(z_{n-1})
  G^{p_{n-1}}_{n-1} 
  \ldots 
  \quad\quad \quad\quad
  \nonumber\\
  \quad\quad\quad
  \ldots
   G^{p_2}_2 
  \Omega_0(z_1)
  G^{p_1}_1 
  \Omega_0(z)
  \rho_\mathrm{S}(t_0)
   \quad\quad 
  \nonumber\\
  \quad\quad \quad\quad 
  \times
  \rb{
    \sum_\mathrm{decomps} \rb{-1}^{N_P} \prod \gamma_{ij} 
  }
  \label{typ_Wick}
 ,
\eeq
where the last factor indicates a sum over all pair decompositions with the relevant Wick sign $(-1)^{N_P}$.
This expansion for $\rho(z)$ can then be resummed to yield the familiar form
\beq
  \rho(z) = \frac{1}{z-{\cal L}_\mathrm{S} -\Sigma(z)}\rho(t_0)
\eeq
with the self-energy
\beq
  \Sigma(z) &=& \sum_{n}^\mathrm{even} \Sigma^{(n)}(z)
  \nonumber\\
  &=& 
  \sum_{n}^\mathrm{even}
   (-i)^n
  \rb{
    \sum_\mathrm{irred.} \rb{-1}^{N_P} \prod \gamma_{ij} 
  }
  \nonumber\\
  &&
  \times
  G^{p_n}_n 
  \Omega_0(z_{n-1})
  G^{p_{n-1}}_{n-1} 
  \ldots 
   G^{p_2}_2 
  \Omega_0(z_1)
  G^{p_1}_1 
  \nonumber
  ,
\eeq
where the sum is over {\it irreducible} contractions only.
Typically one approximates the full kernel by truncating the self energy at a given $n$.  Here we consider only second and fourth order contributions.

At second order, there is only one contraction which reads
\beq
  \Sigma^{(2)}(z) &=& (-1)
  G^{p_2}_2 
  \Omega_0(z_1)
  G^{p_1}_1 \gamma_{21}^{p_2 p_1}
  \label{sig2translate}
  .
\eeq
This translates into diagrams as
\beq
  \Sigma^{(2)}(z) &=& 
  \contraction{}{G}{\underset{z}{-}}{G}
  G \underset{z}{-} G
  .
\eeq
which depicts two system-parts of the tunnel-vertices linked with a bath contraction.  All forefactors are all left implicit in these diagrams.

At fourth order, there are two irreducible contraction schemes: $(41)(32)$ and $(42)(31)$, which we label ``D'' for direct and ``X'' for exchange.  The direct contribution reads
\beq
  \Sigma^{(4D)}(z) &=& 
  G^{p_4}_{\bar{1}}  \Omega_0(z_3)
  G^{p_3}_{\bar{2}}  \Omega_0(z_2)
  \nonumber\\
  &&
  \times
  G^{p_2}_2 \Omega_0(z_1)
  G^{p_1}_1
  \gamma_{41}^{p_4 p_1} \gamma_{32}^{p_3 p_2}
  .
\eeq
and the exchange term is
\beq
  \Sigma^{(4X)}(z) &=& 
  -
  G^{p_4}_{\bar{2}}  \Omega_0(z_3)
  G^{p_3}_{\bar{1}}  \Omega_0(z_2)
  \nonumber\\
  &&
  \times
  G^{p_2}_2  \Omega_0(z_1)
  G^{p_1}_1
   \gamma_{42}^{p_4 p_2} \gamma_{31}^{p_3 p_1}
   .
\eeq
Diagrammatically, we have
\beq
  \Sigma^{(4D)}(z) &=&
  \contraction[2ex]{}{G} {\underset{z}{-} G \underset{z}{-} G \underset{z}{-}} {G}
  \contraction{G \underset{z}{-}} {G} {\underset{z}{-}} {G} 
  G \underset{z}{-} G \underset{z}{-} G \underset{z}{-} G
  \nonumber\\
  \Sigma^{(4X)}(z) &=&
  \contraction {}{G} {\underset{z}{-} G \underset{z}{-}} {G}
  \contraction[2ex]{G\underset{z}{-}} {G} {\underset{z}{-} G \underset{z}{-}} {G}
  G \underset{z}{-} G \underset{z}{-} G\underset{z}{-} G
  .
\eeq
The evaluation of the various integrals required to implement these kernels is discussed in Ref.~\cite{ema09cot}.

Counting is effected by adding counting fields into the tunnel Hamiltonian such that, in  our compact notation, we have
$
 V (p \chi)= \xi_1 a_1 g_1  e^{i p \xi_1 \chi/2}
$.  Correspondingly, the $\chi$-dependent tunnel Liouvilian reads
\beq 
  {\cal L}_V(\chi) &=&-i \xi_1 \sum_p p \sigma^p A^p_1 G^p_1 
    e^{i p s_{\alpha_1}\xi_1 \chi_{\alpha_1}/2}
    \nonumber\\
    &=&-i \xi_1 \sum_p p \sigma^p A^p_1 X^p_1 (\chi)
  \label{AP_LVdef}
\eeq
which defines $X_1^p(\chi) = G^p_1   e^{i p s_{\alpha_1}\xi_1 \chi_{\alpha_1}/2}$ as the symbol for the system-part of of the $\chi$-dependent vertex.

Performing the same expansion as before but with the $\chi$-dependent Liouvillian gives us the $\chi$-dependent nonMarkovian self-energy to fourth order
\beq
  \Sigma(\chi; z) &=& 
  \contraction{}{X}{\underset{z}{-}}{X}
  X \underset{z}{-} X
  +
  \contraction[2ex]{}{X} {\underset{z}{-} X \underset{z}{-} X \underset{z}{-}} {X}
  \contraction{X \underset{z}{-}} {X} {\underset{z}{-}} {X} 
  X \underset{z}{-} X \underset{z}{-} X \underset{z}{-} X
  \nonumber\\
  &&
  +
  \contraction {}{X} {\underset{z}{-} X \underset{z}{-}} {X}
  \contraction[2ex]{X \underset{z}{-}} {X} {\underset{z}{-} X \underset{z}{-}} {X}
  X \underset{z}{-} X \underset{z}{-} X \underset{z}{-} X
  +
  \ldots
  \label{SigXz}
  .
\eeq

\section{Quantum Inhomogeneity}
Writing the quantum MGF of \eq{MGFQfull} as ${\cal G}(\chi,z) = \mathrm{Tr}\left\{\rho(\chi,z)\right\}$ and expanding the propagators we have
\beq
  \rho(\chi,z) 
  &=&
  \lim_{z_0 \to 0} z_0
   \mathrm{Tr}_\mathrm{R}  \left\{ \Pi(\chi;z)\Pi(0;z_0) \varrho(0) \right\}
  \nonumber\\
  &=&  
  \lim_{z_0 \to 0} z_0
  \mathrm{Tr}_\mathrm{R}  
  \left\{
    \rb{
      \Pi_0(z)
      +
      \Pi_0(z){\cal L}_\chi\Pi_0(z)
      +\ldots
    }
    \right.
    \nonumber\\
    &&
    \times
    \left.
    \rb{
      \Pi_0(z_0)
      +
      \Pi_0(z_0){\cal L}_V\Pi_0(z_0)
      +\ldots
    }
    \varrho(0)
  \right\}
  \nonumber
\eeq
We proceed as in the previous section by evaluating contractions between the various tunnel vertices.  There are three possibilities in doing this.  Contractions can be made which gives diagrams which originate solely within the leftmost propagator.  The vertices in such diagrams are all counting vertices and taking all such diagrams together, we obtain the reduced system $\chi$-resolved propagator at finite frequency $z$.  Similarly, diagrams with contractions solely from the rightmost propagator give rise to the plain (i.e. non-counting) system propagator at frequency $z_0\to 0 $.  Finally, contractions can also be made {\it between} the two original propagators.  The sum of all such diagrams can not be resummed and is thus left as a single block and this is what gives rise to the  inhomogeneous term. 
Taking all diagrams into account, the final form of the MGF is
\beq
  {\cal G}(\chi;z)
  &=&
   \lim_{z_0 \to 0} z_0
  \eww{
    \Omega(\chi,z)\rb{\mathbbm{1}+ \Gamma_\mathrm{qm}(\chi,z)} \Omega(0,z_0)
  } 
  \nonumber\\
  &=&
  \eww{
    \Omega(\chi,z)\rb{\mathbbm{1}+ \Gamma_\mathrm{qm}(\chi,z)}
  } 
\eeq
Explicitly writing out only sequential and direct-cotunneling terms, the form for $\Gamma_\mathrm{qm}$ that one arrives at is
\beq
  \Gamma_\mathrm{qm}(\chi;z) &=& 
  \contraction{}{X}{\underset{z}{-}\underset{0}{-}~}{G}
  X \underset{z}{-}\underset{0}{-}~ G
  +
  \contraction[2ex]{}{X} {\underset{z}{-} X \underset{z}{-} X \underset{z}{-}\underset{0}{-} ~} {G}
  \contraction{X \underset{z}{-}} {X} {\underset{z}{-}} {X} 
  X \underset{z}{-} X \underset{z}{-} X \underset{z}{-}\underset{0}{-} ~ G
  \nonumber\\
  &&
  +
  \contraction[2ex]{}{X} {\underset{z}{-} X \underset{z}{-}\underset{0}{-} ~ G \underset{0}{-}} {G}
  \contraction{X \underset{z}{-}} {X} {\underset{z}{-}\underset{0}{-} ~} {G} 
  X \underset{z}{-} X \underset{z}{-}\underset{0}{-} ~ G\underset{0}{-}  G
  +
  \contraction[2ex]{}{X} {\underset{z}{-}\underset{0}{-} ~ G \underset{0}{-} G \underset{0}{-}} {G}
  \contraction{X \underset{z}{-}\underset{0}{-} ~} {G} {\underset{z}{-}} {G} 
  X \underset{z}{-}\underset{0}{-} ~ G \underset{0}{-} G\underset{0}{-}  G
  +\ldots
  \nonumber
  .
\eeq
The exchange-cotunneling diagrams have the same frequency assignments as the three direct diagrams, but with the different pattern of contraction.
The inconvenient thing about this expression is the double free propagators. These can be removed by going one step backwards and noticing that the product of free system-bath propagators can be split as
\beq
  \Pi_0(z)  \Pi_0(z_0) = 
  \frac{1}{z-z_0} \rb{\Pi_0(z_0)-  \Pi_0(z) } 
  .
\eeq
Performing this separation and  re-doing the contractions, we see that we can make the diagrammatic replacement
\beq
  \underset{z}{-}\underset{0}{-} =  
  \frac{1}{z-z_0} \rb{\underset{0}{-} + (-1)  \underset{z}{-} } 
\eeq
The inhomogeneity block becomes
\begin{widetext}
\beq
  z\Gamma_\mathrm{qm}(\chi;z) &=&  
  \contraction{}{X}{\underset{0}{-}}{G}
  X \underset{0}{-} G
  + (-1)
  \contraction{}{X}{\underset{z}{-}}{G}
  X \underset{z}{-} G
  +
  \contraction[2ex]{}{X} {\underset{z}{-} X \underset{z}{-} X \underset{0}{-}} {G}
  \contraction{X \underset{z}{-}} {X} {\underset{z}{-}} {X} 
  X \underset{z}{-} X \underset{z}{-} X \underset{0}{-} G
  + (-1)
  \contraction[2ex]{}{X} {\underset{z}{-} X \underset{z}{-} X \underset{z}{-}} {G}
  \contraction{X \underset{z}{-}} {X} {\underset{z}{-}} {X} 
  X \underset{z}{-} X \underset{z}{-} X \underset{z}{-} G
  \nonumber\\
  &&
  +
  \contraction[2ex]{}{X} {\underset{z}{-} X \underset{0}{-} G \underset{0}{-}} {G}
  \contraction{X \underset{z}{-}} {X} {\underset{0}{-}} {G} 
  X \underset{z}{-} X \underset{0}{-} G \underset{0}{-} G
  + (-1)
  \contraction[2ex]{}{X} {\underset{z}{-} X \underset{z}{-} G \underset{0}{-}} {G}
  \contraction{X \underset{z}{-}} {X} {\underset{z}{-}} {G} 
  X \underset{z}{-} X \underset{z}{-} G \underset{0}{-} G
  +
  \contraction[2ex]{}{X} {\underset{0}{-} G \underset{0}{-} G \underset{0}{-}} {G}
  \contraction{X \underset{0}{-}} {G} {\underset{0}{-}} {G} 
  X \underset{0}{-} G \underset{0}{-} G \underset{0}{-} G
  + (-1)
  \contraction[2ex]{}{X} {\underset{z}{-} G \underset{0}{-} G \underset{0}{-}} {G}
  \contraction{X \underset{z}{-}} {G} {\underset{0}{-}} {G} 
  X \underset{z}{-} G \underset{0}{-} G \underset{0}{-} G
  +\ldots
  \nonumber
  .
\eeq
Comparison of this form with \eq{SigXz} allows us to write
\beq
  z\Gamma_\mathrm{qm}(\chi;z) &=&  
  - \Sigma(\chi;z)
  +
  \contraction{}{X}{\underset{0}{-}}{G}
  X \underset{0}{-} G
  +
  \contraction{}{X}{\underset{z}{-}}{\Delta}
  X \underset{z}{-} \Delta
  \nonumber\\
  &&
  +
  \contraction[2ex]{}{X} {\underset{0}{-} G \underset{0}{-} G \underset{0}{-}} {G}
  \contraction{X \underset{0}{-}} {G} {\underset{0}{-}} {G} 
  X \underset{0}{-} G \underset{0}{-} G \underset{0}{-} G
  +
  \contraction[2ex]{}{X} {\underset{z}{-} \Delta \underset{0}{-} G \underset{0}{-}} {G}
  \contraction{X \underset{z}{-}} {\Delta} {\underset{0}{-}} {G} 
  X \underset{z}{-} \Delta \underset{0}{-} G \underset{0}{-} G
  +
  \contraction[2ex]{}{X} {\underset{z}{-} X \underset{z}{-} \Delta \underset{0}{-}} {G}
  \contraction{X \underset{z}{-}} {X} {\underset{z}{-}} {\Delta} 
  X \underset{z}{-} X \underset{z}{-} \Delta \underset{0}{-} G
  +
  \contraction[2ex]{}{X} {\underset{z}{-} X \underset{z}{-} X \underset{z}{-}} {\Delta}
  \contraction{X \underset{z}{-}} {X} {\underset{z}{-}} {X} 
  X \underset{z}{-} X \underset{z}{-} X \underset{z}{-} \Delta
  +\ldots
  \label{Gamma}
\eeq
where we have introduced the shorthand
\beq
  \Delta = X + (-1) G.
\eeq  
This final form for the inhomogeneity is rather useful in evaluating the cumulants, because any $\Delta$ that remains undifferentiated gives zero when the $\chi \to 0$ limit is taken.

\section{Current blocks}
Writing out explicitly only the sequential and direct-cotunneling contributions, the first and second current blocks read
\beq
  {\cal J}^{(1)}(z_1,z_0) &=& 
  \contraction{}{G}{\underset{z_1}{-}}{X'}
  G \underset{z_1}{-} X'
  +
   \contraction{}{X'}{\underset{z_0}{-}}{G}
  X' \underset{z_0}{-} G
  \nonumber\\
  &&
  +
  \contraction[2ex]{}{X'} {\underset{z_0}{-} G \underset{z_0}{-} G \underset{z_0}{-}} {G}
  \contraction{X' \underset{z_0}{-}} {G} {\underset{z_0}{-}} {G} 
  X' \underset{z_0}{-} G \underset{z_0}{-} G \underset{z_0}{-} G
 +
  \contraction[2ex]{}{G} {\underset{z_1}{-} X' \underset{z_0}{-} G \underset{z_0}{-}} {G}
  \contraction{G \underset{z_1}{-}} {X'} {\underset{z_0}{-}} {G} 
  G \underset{z_1}{-} X' \underset{z_0}{-} G \underset{z_0}{-} G
  +
  \contraction[2ex]{}{G} {\underset{z_1}{-} G \underset{z_1}{-} X' \underset{z_0}{-}} {G}
  \contraction{G \underset{z_1}{-}} {G} {\underset{z_1}{-}} {X'} 
  G \underset{z_1}{-} G \underset{z_1}{-} X' \underset{z_0}{-} G
  +
  \contraction[2ex]{}{G} {\underset{z_1}{-} G \underset{z_1}{-} G \underset{z_1}{-}} {X'}
  \contraction{G \underset{z_1}{-}} {G} {\underset{z_1}{-}} {G} 
  G \underset{z_1}{-} G \underset{z_1}{-} G \underset{z_1}{-} X'
  +
  \ldots
\eeq
%
%
\beq
  \textstyle{ \frac{1}{2}} {\cal J}_{\gamma_2\gamma_1}^{(2)}(z_2,z_1,z_0) 
  &=& 
  \contraction{}{X'}{\underset{z_1}{-}}{X'}
  X' \underset{z_1}{-} X'
  +  
  \textstyle{ \frac{1}{2}}
  \contraction{}{X''}{\underset{z_0}{-}}{G}
  X'' \underset{z_0}{-} G
  +
  \textstyle{ \frac{1}{2}}
  \contraction{}{G}{\underset{z_2}{-}}{X''}
  G \underset{z_2}{-} X''
  \nonumber\\
  &&
  +
  \contraction[2ex]{}{X'} {\underset{z_1}{-} X' \underset{z_0}{-} G \underset{z_0}{-}} {G}
  \contraction{X' \underset{z_1}{-}} {X'} {\underset{z_0}{-}} {G} 
  X' \underset{z_1}{-} X' \underset{z_0}{-} G \underset{z_0}{-} G
  +
  \contraction[2ex]{}{X'} {\underset{z_1}{-} G \underset{z_1}{-} X' \underset{z_0}{-}} {G}
  \contraction{X' \underset{z_1}{-}} {G} {\underset{z_1}{-}} {X'} 
  X' \underset{z_1}{-} G \underset{z_1}{-} X' \underset{z_0}{-} G
  +
  \contraction[2ex]{}{X'} {\underset{z_1}{-} G \underset{z_1}{-} G \underset{z_1}{-}} {X'}
  \contraction{X' \underset{z_1}{-}} {G} {\underset{z_1}{-}} {G} 
  X' \underset{z_1}{-} G \underset{z_1}{-} G \underset{z_1}{-} X'
  \nonumber\\
  &&
  +
  \contraction[2ex]{}{G} {\underset{z_2}{-} X' \underset{z_1}{-} X' \underset{z_0}{-}} {G}
  \contraction{G \underset{z_2}{-}} {X'} {\underset{z_1}{-}} {X'} 
  G \underset{z_2}{-} X' \underset{z_1}{-} X' \underset{z_0}{-} G
  +
  \contraction[2ex]{}{G} {\underset{z_2}{-} X' \underset{z_1}{-} G \underset{z_1}{-}} {X'}
  \contraction{G \underset{z_2}{-}} {X'} {\underset{z_1}{-}} {G} 
  G \underset{z_2}{-} X' \underset{z_1}{-} G \underset{z_1}{-} X'
  +
  \contraction[2ex]{}{G} {\underset{z_2}{-} G \underset{z_2}{-} X' \underset{z_1}{-}} {X'}
  \contraction{G \underset{z_2}{-}} {G} {\underset{z_2}{-}} {X'} 
  G \underset{z_2}{-} G \underset{z_2}{-} X' \underset{z_1}{-} X'
  \nonumber\\
  &&
  +
  \textstyle{ \frac{1}{2}}
  \contraction[2ex]{}{X''} {\underset{z_0}{-} G \underset{z_0}{-} G \underset{z_0}{-}} {G}
  \contraction{X'' \underset{z_0}{-}} {G} {\underset{z_0}{-}} {G} 
  X'' \underset{z_0}{-} G \underset{z_0}{-} G \underset{z_0}{-} G
  +
  \textstyle{ \frac{1}{2}}
  \contraction[2ex]{}{G} {\underset{z_2}{-} X'' \underset{z_0}{-} G \underset{z_0}{-}} {G}
  \contraction{G \underset{z_2}{-}} {X''} {\underset{z_0}{-}} {G} 
  G \underset{z_2}{-} X'' \underset{z_0}{-} G \underset{z_0}{-} G
  +
  \textstyle{ \frac{1}{2}}
  \contraction[2ex]{}{G} {\underset{z_2}{-} G \underset{z_2}{-} X'' \underset{z_0}{-}} {G}
  \contraction{G \underset{z_2}{-}} {G} {\underset{z_2}{-}} {X''} 
  G \underset{z_2}{-} G \underset{z_2}{-} X'' \underset{z_0}{-} G
  +
  \textstyle{ \frac{1}{2}}
  \contraction[2ex]{}{G} {\underset{z_2}{-} G \underset{z_2}{-} G \underset{z_2}{-}} {X''}
  \contraction{G \underset{z_2}{-}} {G} {\underset{z_2}{-}} {G} 
  G \underset{z_2}{-} G \underset{z_2}{-} G \underset{z_2}{-} X''
  +
  \ldots
\eeq
\end{widetext}




\begin{thebibliography}{99}
\bibitem{LMGFN} Diagrams which have leftmost a plain tunnel vertex, $G$ disappear in the expectation value since the sum over the Keldysh index of a leftmost tunnel vertex yields two contributions which, under the trace, are identical but with opposite sign.  A particular example is the expectation value of the self-energy $\eww{\Sigma(\chi=0)}=0$ since $\Sigma(\chi=0)$ contains only tunnel vertices.
\bibitem{interpFN}  It should be stressed that these difficulties of interpretation arise from the attempt to shoe-horn quantum mechanical objects (the MGF; $\chi$-resolved NMME) into classical language $P(n;t)$.  No problems arise in the MGF is just used as a method of calculating the various moments.
\end{thebibliography}
\end{document}